\documentclass[%
 aip,
 jmp,%
 amsmath,amssymb,
 reprint,%
]{revtex4-1}

\usepackage{graphicx}% Include figure files
\usepackage{dcolumn}% Align table columns on decimal point
\usepackage{bm}% bold math
\begin{document}

\preprint{AIP/123-QED}

\title[Strange Attractors Characterizing the Osmotic Instability]{Strange Attractors Characterizing the Osmotic Instability\footnote{Error!}}
% Force line breaks with \\
% \thanks{Footnote to title of article.}

\author{Stephan I. Tzenov}

\affiliation{Creative Way Science at Plovdiv University, Plovdiv, Bulgaria}

\altaffiliation{Creative Way Science home page: http://www.creativeway.eu/}


\email{stephan.tzenov@gmail.com}

% \affiliation{
% Authors' institution and/or address%\\This line break forced with \textbackslash\textbackslash
% }%

%\author{Kiril B. Marinov}

%\affiliation{ASTeC, STFC Daresbury Laboratory, Keckwick Lane, Daresbury, WA4 4AD, United Kingdom.}

%\altaffiliation{The Cockcroft Institute, Keckwick Lane, Daresbury, WA4 4AD, United Kingdom.}

% \author{C. Author}
% \homepage{http://www.Second.institution.edu/~Charlie.Author.}
% \affiliation{%
% Second institution and/or address%\\This line break forced% with \\
% }%

\date{\today}% It is always \today, today,

\begin{abstract}
In the present paper a simple dynamical model for computing the osmotically driven fluid flow in a variety of complex, non equilibrium situations is derived from first principles. Using the Oberbeck-Boussinesq approximation, the basic equations describing the process of forward osmosis have been obtained. It has been shown that these equations are very similar to the ones used to model the free Rayleigh-Benard convection. The difference is that while in the case of thermal convection the volume expansion is driven by the coefficient of thermal expansion, the key role for the osmotic instability is played by the coefficient of isothermal compressibility $\beta$. In addition, it has been shown that the osmotic process represents a propagation of standing waves with time-dependent amplitudes and phase velocity, which equals the current velocity of the solvent passing through the semi-permeable membrane. The evolution of the amplitudes of the osmotic waves is exactly following the dynamics of a strange attractor of Lorenz type with corresponding control parameters.

\end{abstract}

\pacs{47.10.A, 47.52.+j, 47.56.+r}% PACS, the Physics and Astronomy
\keywords{Forward Osmosis, Fluid Mixture, Strange Attractor}%Use showkeys class
                              %option if keyword
\maketitle

% \begin{quotation}
% The ``lead paragraph'' is encapsulated with the \LaTeX\
% \verb+quotation+ environment and is formatted as a single paragraph before the first section heading.
% (The \verb+quotation+ environment reverts to its usual meaning after the first sectioning command.)
% Note that numbered references are allowed in the lead paragraph.
%
% The lead paragraph will only be found in an article being prepared for the journal \textit{Chaos}.
% \end{quotation}

\section{\label{sec:intro}Introduction}

Osmotic effects are thought to play an important role in many physical systems associated with biology, medicine and modern technological applications. Typical examples of technological applications include microscopic devices designed to pump fluids, actuate forces through swelling, or deliver drug doses. Some macroscopic osmotic mechanisms in biology include the exchange of blood constituents in capillaries with surrounding tissues and the processing of fluids in tissues of layered epithelial cells in the intestines and kidneys.

Lately, forward osmosis is receiving continuously increasing interest as an alternative mechanism to the conventional hydrostatic pressure-driven membrane processes. Compared with pressure-driven systems, the forward osmosis is a technique driven by an osmotic pressure gradient between fluid compounds separated by a semi-permeable membrane. As a consequence, it does not require hydraulic pressure operation and is potentially more cost-effective compared with other methods such as reverse osmosis. In addition, forward osmosis has a lower propensity to membrane fouling, possibly due to the absence of hydraulic pressure

The classical thermodynamic description of osmotic pressure is sometimes quite sufficient to describe equilibrium properties and compute most quantities of interest without reference to any molecular model \cite{Finkelstein}. It is however intuitively clear that certain molecular models and a microscopic picture underly the equilibrium situations and can be quite useful when thinking about osmotic flow in unfamiliar situations. These equilibrium situations are well described in the classical literature \cite{Hill,Reichl}. However, osmotic flows are generally handled by employing equilibrium quantities to describe non equilibrium processes.

Recently, the methods of fluid dynamics have become increasingly popular for the analysis of flow patterns and complex coherent structures emerging in membrane systems. Many different hydrodynamic models treating pressure-driven membrane systems have been presented during the years. While some of them focus on the effect of various membrane properties like solute rejection and solution properties such as density, viscosity and diffusivity, other models stress their attention on mass transfer optimization by changing the cell geometry in order to achieve an optimal flow by using eddy-promoting spacers. Unfortunately, the physical nature of osmotic processes is given scant attention in the literature up to date.

The present paper is aimed at filling the above mentioned gap in the existing literature. In Sections \ref{sec:generalities} and \ref{sec:hydro} we derive from first principles a quite general hydrodynamic model for the description of forward osmosis, which is similar to the model presented in the paper by M.F. Gruber et al \cite{Gruber}. Using the Oberbeck-Boussinesq approximation we derive in Section \ref{sec:linear} the basic equations describing the process of forward osmosis, which are very similar to the ones used by Saltzman \cite{Saltzman} to model the free Rayleigh-Benard convection. In Section \ref{sec:attractor} we show that the osmotic process represents a propagation of standing waves with time-dependent amplitudes, the evolution of which is exactly following a strange attractor with corresponding control parameters. Finally, Section \ref{sec:remarks} is dedicated to conclusions and outlook.

\section{\label{sec:generalities}Formulation of the problem and basic equations}

To be as general as possible, let us consider a system consisting of $N$ particles, which occupies volume $V$ in the configuration space. The statistical mechanical description of complex systems contains in general two fundamental time scales. The first one is the minimum {\it correlation time}, which characterizes the individual inter-particle (binary, ternary, etc.) interactions, both short-range and long-range ones. The second time scale called the {\it relaxation time} is typical for the description of (relatively) long-term behaviour of many-particle systems, on which the rich variety of mesoscopic and macroscopic features and structures build up in the process of evolution towards equilibrium. As a rule, the relaxation time is much greater than the correlation time. In between these two extremal scales, there exists yet a third characteristic time scale (the so called {\it physically infinitesimal time} scale $t_{ph}$) corresponding to an intermediate coarse-grained description \cite{Klimontovich,TzenovBook} of the relaxation dynamics.

The generalized Boltzmann equation for the one-particle distribution function $f {\left( {\bf x}, {\bf p}; t \right)}$ can be written as \cite{TzenovBook}
\begin{equation}
\partial_t f + {\bf v} \cdot {\bf \nabla}_x f + {\bf \nabla}_p \cdot {\left[ {\left( {\bf F} + {\bf F}_c \right)} f  \right]} = {\cal J}_{col} {\left( f  \right)} + {\frac {1} {t_{ph}}} {\left( {\widetilde{f}} - f \right)}, \label{Boltzmann}
\end{equation}
\noindent where ${\bf F}_c$ is the average interaction force between particles comprising the system under consideration, ${\bf F}$ is some external force applied on the system, and ${\cal J}_{col}$ is the well-known Boltzmann collision integral. The function ${\widetilde{f}} {\left( {\bf x}, {\bf p}; t \right)}$ represents the distribution function $f$ smoothed over the infinitesimal volume $l_{ph}^3$.  It can be shown \cite{TzenovBook,Capillary} that the additional collision term in the generalized kinetic equation (\ref{Boltzmann}) can be cast in the form of a Fokker-Planck operator acting on the distribution function $f {\left( {\bf x}, {\bf p}; t \right)}$, where the corresponding drift vector ${\bf A}$ and diffusion matrix $B_{mn}$ are functions of the coordinate ${\bf x}$ in general. Following the standard procedure of building the conservation laws (equivalent to calculating the hydrodynamic moments) up to second order \cite{Capillary,Huang}, one arrives at the following system of equations
\begin{equation}
\partial_t \varrho + {\bf \nabla} \cdot {\left[ \varrho {\left( {\bf v} + {\bf A} \right)} \right]} = {\frac {1} {2}} \nabla_k \nabla_l {\left( B_{kl} \varrho \right)}, \label{Contin}
\end{equation}
\begin{eqnarray}
\partial_t {\left( \varrho v_k \right)} + \nabla_l {\left( \varrho v_k v_l \right)} = {\frac {\varrho F_k} {m}} - {\bf \nabla} \cdot {\left( {\bf A} \varrho v_k \right)} \nonumber
\end{eqnarray}
\begin{equation}
- \nabla_l {\cal P}_{kl} + {\frac {1} {2}} \nabla_l \nabla_n {\left( B_{ln} \varrho v_k \right)}, \label{MomBalanc}
\end{equation}
\noindent where ${\bf A}$ is a drift vector associated with the external force, and $B_{mn}$ is a diffusion matrix. In the equations above, the well-known hydrodynamic quantities, such as the mass density $\varrho$, the mean current velocity ${\bf v}$ of a fluid element and the hydrodynamic stress tensor $\Pi_{kl}$ are defined as
\begin{equation}
\varrho {\left( {\bf x}; t \right)} = m n \int {\rm d}^3 {\bf p} f {\left( {\bf x}, {\bf p}; t \right)}, \label{MassDense}
\end{equation}
\begin{equation}
\varrho {\left( {\bf x}; t \right)} {\bf v} {\left( {\bf x}; t \right)} = n \int {\rm d}^3 {\bf p} {\bf p} f {\left( {\bf x}, {\bf p}; t \right)}, \label{ForwardVel}
\end{equation}
\begin{equation}
\Pi_{kl} {\left( {\bf x}; t \right)} = {\frac {n} {m}} \int {\rm d}^3 {\bf p} p_k p_l f {\left( {\bf x}, {\bf p}; t \right)} =  \varrho v_k v_l + {\cal P}_{kl}, \label{StressTen}
\end{equation}
\noindent where the thermodynamic limit $n = \lim_{N,V \to \infty} {\left( N / V \right)}$ has been taken.

To proceed further, we assume that there are no intrinsic instabilities which develop in the course of the solvent's flow, however the dynamics of the solute particles regarded as well mixed tracers are well described by Eq.(\ref{Contin}). Since the solute and the solvent are miscible, the density of the solution will be equal to the sum of the densities of the solute and the solvent, while its current velocity will be simply equal to the current density of the solvent. Simple manipulation of Eqs. (\ref{Contin}) and (\ref{MomBalanc}) yields a system of hydrodynamic equations \cite{Capillary}, which will be the starting point of the subsequent analysis. Writing the latter in explicit form, we obtain
\begin{equation}
\partial_t \varrho + {\bf \nabla} \cdot {\left( \varrho {\bf v} \right)} = 0, \label{Continuity}
\end{equation}
\begin{equation}
\partial_t {\left( \varrho {\cal C} \right)} + {\bf \nabla} \cdot {\left[ \varrho {\cal C} {\left( {\bf v} - {\bf u} \right)} \right]} = 0, \label{ContinuityS}
\end{equation}
\begin{equation}
u_k = {\frac {B_{kl}} {2 \varrho {\cal C}}} \nabla_l {\left( \varrho {\cal C} \right)}, \label{Fick}
\end{equation}
\begin{equation}
\partial_t {\bf v} + {\bf v} \cdot {\bf \nabla} {\bf v} = {\frac {\bf F} {M}} - {\frac {1} {\varrho_s}} {\bf \nabla} {\cal P} + {\frac {1} {\varrho_s}} {\left[ {\frac {3 \zeta + \eta} {3}} {\bf \nabla} {\left( {\bf \nabla} \cdot {\bf v} \right)} + \eta {\nabla}^2 {\bf v} \right]}. \label{NavierStokes}
\end{equation}

In the equations above $\varrho$ is the mass density of the solvent, $\varrho_s$ is the mass density of the solution and ${\bf v}$ is their current velocity. Furthermore, by $m$ we denote the molecular mass of the solvent and by $m_s$ the molecular mass of the solute, such that the molecular mass of the solution is $M = m + m_s$. The quantity $\mu = m / m_s$ is the mass fraction and ${\cal C} = \varrho_{sol} / \varrho$ is the mass concentration of the solution, where $\varrho_{sol}$ is the mass density of the solute. Obviously, $\varrho_s = \varrho + \varrho_{sol} = \varrho {\left( 1 + {\cal C} \right)}$. In addition, $\eta$ is the well-known coefficient of dynamic viscosity and $\zeta$ is the second viscosity coefficient.

Equations (\ref{Continuity}) and (\ref{ContinuityS}) are the well-known continuity equations describing the conservation law of mass transport for the solvent and for the solute, respectively, while Eq. (\ref{Fick}) is the first Fick's law relating the diffusive flux to the gradient of the mass density. Equation (\ref{NavierStokes}) is the classical Navier-Stokes equation describing the momentum balance in the solution as a whole. The pressure term ${\cal P}$ is a sum of the hydrodynamic pressure (in the bulk of the solution)
\begin{equation}
{\cal P} = {\frac {k_B T} {M}} \varrho, \label{Pressure}
\end{equation}
\noindent the osmotic pressure acting on the solvent on one side of the semi-permeable membrane and additional pressure caused by possible external forces acting on the system. According to van't Hoff's law, the osmotic pressure is given by
\begin{equation}
{\cal P}_o = k_B n_A T \ln {\left( 1 + \mu {\cal C} \right)}, \label{OsmPressure}
\end{equation}
\noindent where $k_B$ is the Boltzmann constant, $T$ is the temperature, and $n_A = N_A / V$ is the Avogadro number density related to the volume $V$.

\section{\label{sec:hydro}Non-dimensional form of the fluid dynamic equations}

In general, the diffusion matrix $B_{mn}$ depends not only on the coordinate ${\bf x}$ but on the mass concentration ${\cal C}$ as well. In what follows we consider the simplest case, where the diffusion matrix is isotropic and constant
\begin{equation}
B_{kl} = B \delta_{kl}. \label{Diffusion}
\end{equation}
\noindent It is convenient to cast the basic hydrodynamic equations introduced in the preceding Section in a non-dimensional form. We introduce the non-dimensional time, coordinate and velocity according to
\begin{equation}
t^{\ast} = {\frac {\omega_c t} {2 \pi}}, \qquad {\bf x}^{\ast} = {\frac {\bf x} {L_c}}, \qquad {\bf v}^{\ast} = {\frac {\bf v} {V_c}}, \qquad {\bf u}^{\ast} = {\frac {\bf u} {V_c}}, \label{Nondimen}
\end{equation}
\noindent and the non-dimensional mass density and pressure
\begin{equation}
\varrho^{\ast} = {\frac {\varrho} {\varrho_c}}, \qquad \qquad {\cal P}^{\ast} = {\frac {\cal P} {\varrho_c V_c^2}}, \label{Mass}
\end{equation}
\noindent where $\varrho_c$ is an arbitrarily chosen constant reference mass density, while $\omega_c$, $L_c$ and $V_c$ are some characteristic frequency, length and speed. Then, the hydrodynamic equations (\ref{Continuity}) - (\ref{NavierStokes}) can be written as
\begin{equation}
\partial_t^{\ast} \varrho^{\ast} + {\bf \nabla}^{\ast} \cdot {\left( \varrho^{\ast} {\bf v}^{\ast} \right)} = 0, \label{Continuit}
\end{equation}
\begin{equation}
\partial_t^{\ast} {\left( \varrho^{\ast} {\cal C} \right)} + {\bf \nabla}^{\ast} \cdot {\left[ \varrho^{\ast} {\cal C} {\left( {\bf v}^{\ast} - {\bf u}^{\ast} \right)} \right]} = 0, \label{ContinuitS}
\end{equation}
\begin{equation}
{\bf u}^{\ast} = {\frac {\cal D} {2 \varrho^{\ast} {\cal C}}} {\bf \nabla}^{\ast} {\left( \varrho^{\ast} {\cal C} \right)}, \label{Ficknd}
\end{equation}
\begin{eqnarray}
\partial_t^{\ast} {\bf v}^{\ast} + {\bf v}^{\ast} \cdot {\bf \nabla}^{\ast} {\bf v}^{\ast} = - {\bf \nabla}^{\ast} \Phi - {\frac {n_A \Theta} {n_c \varrho_s^{\ast}}} {\bf \nabla}^{\ast} {\left[ \ln {\left( 1 + \mu {\cal C} \right)} \right]}  \nonumber
\end{eqnarray}
\begin{equation}
- {\frac {1} {\varrho_s^{\ast}}} {\bf \nabla}^{\ast} {\cal P}^{\ast} + {\frac {1} {R_{e1}}} {\bf \nabla}^{\ast} {\left( {\bf \nabla}^{\ast} \cdot {\bf v}^{\ast} \right)} + {\frac {1} {R_e}} {\nabla}^{\ast 2} {\bf v}^{\ast}, \label{NavierStoke}
\end{equation}

In the above system of equations (\ref{Continuit}) - (\ref{NavierStoke}) the following notations have been introduced. The quantity $\Phi = \varphi / {\left( m V_c^2 \right)}$ is the dimensionless external potential ${\left( L_c {\bf F} = - {\bf \nabla}^{\ast} \varphi \right)}$ acting on the system, while the quantity $\Theta = k_B T / {\left( m V_c^2 \right)}$ measures the dimensionless thermal energy. In addition
\begin{equation}
{\cal D} = {\frac {B} {L_c V_c}}, \qquad \varrho_c = m n_c, \label{Variables}
\end{equation}
\noindent while $R_e$ is the well-known Reynolds number
\begin{equation}
R_{e1} = {\frac {\varrho_s L_c V_c} {\zeta + \eta / 3}}, \qquad R_e = {\frac {\varrho_s L_c V_c} {\eta}}. \label{Reynolds}
\end{equation}
\noindent It is worthwhile to note that for the case of incompressible solvent the fourth term on the right-hand-side of Eq. (\ref{NavierStoke}) disappears. For the sake of notational simplicity of the forthcoming exposition, the star superscript over the dimensionless variables will be omitted.

\section{\label{sec:linear}Derivation of the basic model of forward osmosis}

In the present Section we analyse the case of incompressible solvent ${\left( {\bf {\nabla}} \cdot {\bf v} = 0 \right)}$. Due to the action of the osmotic force, the solution as a whole can be compressed/diluted as will be shown in the subsequent exposition. Let us also assume that the only external force acting on the system is the force of gravity. Thus, hydrodynamic equations can be rewritten as follows
\begin{equation}
\partial_t \varrho + {\bf v} \cdot {\bf \nabla} \varrho = 0, \label{Continui}
\end{equation}
\begin{equation}
\partial_t {\cal C} + {\bf v} \cdot {\bf \nabla} {\cal C} = {\frac {\cal D} {2 \varrho}} {\bf \nabla}^2 {\left( \varrho {\cal C} \right)}, \label{ContinS}
\end{equation}
\begin{equation}
\partial_t {\bf v} + {\bf v} \cdot {\bf \nabla} {\bf v} = - {\bf \nabla} \Phi - {\frac {1} {\varrho_s}} {\bf \nabla} {\cal P}_t + {\frac {1} {R_e}} {\bf \nabla}^2 {\bf v}. \label{NavierStok}
\end{equation}
\noindent Here
\begin{equation}
{\bf \nabla} \Phi = {\left( 0, 0, g_0 \right)}, \qquad \qquad g_0 = {\frac {g L_c} {V_c^2}}, \label{Gravity}
\end{equation}
\noindent with $g$ being the standard gravity constant, and
\begin{equation}
{\cal P}_t = {\cal P} + {\frac {n_A \Theta} {n_c}} \ln {\left( 1 + \mu {\cal C} \right)}, \label{TotPress}
\end{equation}
\noindent is the total pressure applied on the solution.

Since the vertical component of the current velocity vanishes ${\left( v_z = 0 \right)}$ in the quiescent fluid solution outside the boundary layer, the momentum balance equation becomes
\begin{equation}
\partial_z {\cal P}_{\infty} = - g_0 \varrho_{s \infty} - {\frac {n_A \Theta \mu} {n_c {\left( 1 + \mu {\cal C}_{\infty} \right)}}} \partial_z {\cal C}_{\infty}. \label{Quiescez}
\end{equation}
\noindent Noting that $v_x \approx {\rm const}$ in the boundary layer, the balance of the momentum in the $x$-direction yields
\begin{equation}
{\cal P} = G {\left( z \right)} - {\frac {n_A \Theta} {n_c}} \ln {\left( 1 + \mu {\cal C} \right)}, \label{Quiescex}
\end{equation}
\noindent where to this end $G {\left( z \right)}$ is an arbitrary function of the vertical coordinate $z$. Matching the pressure in the region of quiescent fluid, from the above expressions we readily obtain
\begin{equation}
\partial_z {\cal P} = \partial_z {\cal P}_{\infty} = - g_0 \varrho_{s \infty} - {\frac {n_A \Theta \mu} {n_c {\left( 1 + \mu {\cal C}_{\infty} \right)}}} \partial_z {\cal C}_{\infty}. \label{Quiescez1}
\end{equation}
\noindent In addition, it follows also that $\partial_z G = - g_0 \varrho_{s \infty}$.

Let us further introduce the coefficient of isothermal compressibility of the solution
\begin{eqnarray}
\beta = - {\frac {1} {V}} {\left( {\frac {\partial V} {\partial {\cal P}}} \right)}_{\Theta, N = {\rm const}} = {\frac {1} {\varrho_s}} {\left( {\frac {\partial \varrho_s} {\partial {\cal P}}} \right)}_{\Theta = {\rm const}} \nonumber
\end{eqnarray}
\begin{equation}
= {\frac {1} {1 + {\cal C}}} {\left( {\frac {\partial {\cal C}} {\partial {\cal P}}} \right)}_{\Theta = {\rm const}}, \label{Compressibility}
\end{equation}
\noindent or approximately
\begin{equation}
\beta \approx {\frac {1} {1 + {\cal C}}} {\frac {\Delta {\cal C}} {\Delta {\cal P}}} = {\frac {1} {1 + {\cal C}}} {\frac {{\cal C}_{\infty} - {\cal C}} {{\cal P}_{\infty} - {\cal P}}}. \label{Compressibility1}
\end{equation}
\noindent Using the above considerations and taking into account only terms linear with respect to the mass fraction $\mu$, we finally arrive at
\begin{equation}
\partial_t c + {\bf v} \cdot {\bf \nabla} c = {\frac {\cal D} {2}} {\bf \nabla}^2 c, \label{ContinSL}
\end{equation}
\begin{equation}
\partial_t v_x + {\bf v} \cdot {\bf \nabla} v_x = - {\frac {\lambda \Theta} {\varrho_s}} \partial_x c + {\frac {1} {R_e}} {\bf \nabla}^2 v_x, \label{NavierStokxL}
\end{equation}
\begin{equation}
\partial_t v_z + {\bf v} \cdot {\bf \nabla} v_z = \lambda \Theta g_0 \beta c - {\frac {\lambda \Theta} {\varrho_s}} \partial_z c + {\frac {1} {R_e}} {\bf \nabla}^2 v_z, \label{NavierStokzL}
\end{equation}
\noindent where
\begin{equation}
\lambda = {\frac {n_A \mu} {n_c}}, \label{Lambda}
\end{equation}
\noindent and $c = {\cal C} - {\cal C}_{\infty}$.

Incompressibility of the solvent implies that
\begin{equation}
v_x = \partial_z \Psi = \Psi_z, \qquad \qquad v_z = - \partial_x \Psi = - \Psi_x, \label{Incompress}
\end{equation}
\noindent where $\Psi {\left( x, z \right)}$ is an arbitrary function of its arguments. The boundary condition at the semi-permeable membrane located at $x=0$ can be written as \cite{Membrane}
\begin{equation}
v_{x0} {\left( z; t \right)} = {\frac {A} {1 - A}} {\sqrt{\frac {\Theta} {2 \pi}}} {\cal C} {\left( 0, z; t \right)}, \label{Boundary}
\end{equation}
\noindent where $A$ is the ratio of the total surface of all holes to the membrane surface. The average velocity of the fluid current passing through the membrane for a certain time lapse $T$ is
\begin{eqnarray}
v_0 = {\frac {A} {z_M T {\left( 1 - A \right)}}} {\sqrt{\frac {\Theta} {2 \pi}}} \int \limits_0^T {\rm d} t \int \limits_0^{z_M} {\rm d} z {\cal C} {\left( 0, z, t \right)} \nonumber
\end{eqnarray}
\begin{equation}
\approx {\frac {A} {1 - A}} {\sqrt{\frac {\Theta} {2 \pi}}} {\cal C}_{\infty}, \label{Boundary1}
\end{equation}
\noindent where $z_M = L_M / L_c$ is the transverse dimension of the membrane. Noting that the system of equations (\ref{ContinSL}) - (\ref{NavierStokzL}) is invariant under Galilean transformation $v_x \longrightarrow v_x + v_0$, we can incorporate the boundary condition (\ref{Boundary1}) by a simple change of the longitudinal variable
\begin{equation}
s = x - v_0 t. \label{Galilei}
\end{equation}
\noindent Simple and obvious manipulation of equations (\ref{ContinSL}) - (\ref{NavierStokzL}) with due account of Eq. (\ref{Incompress}) yields
\begin{equation}
\partial_t c + \Psi_z \partial_s c - \Psi_s \partial_z c = {\frac {\cal D} {2}} {\bf \nabla}^2 c, \label{ContinSLb}
\end{equation}
\begin{equation}
\partial_t {\bf \nabla}^2 \Psi + \Psi_z {\bf \nabla}^2 \Psi_s - \Psi_s {\bf \nabla}^2 \Psi_z = - \lambda \Theta g_0 \beta \partial_s c + {\frac {1} {R_e}} {\bf \nabla}^2 {\bf \nabla}^2 \Psi. \label{Hydropot}
\end{equation}
\noindent The above system of equations for the unknown functions $\Psi$ and $c$ comprise the basic set of equations for the description of the process of forward osmosis. It will be our starting point to derive the equations of the strange attractor in the next section.

\section{\label{sec:attractor}Derivation and stability analysis of the strange attractor}

Let us introduce the deviation $\Sigma {\left( s, z; t \right)}$ from the linear profile of the mass concentration of the solution
\begin{equation}
c {\left( s, z; t \right)} = c_0 - \Delta c z + \Sigma {\left( s, z; t \right)}. \label{LinearProf}
\end{equation}
\noindent The basic equations (\ref{ContinSLb}) and (\ref{Hydropot}) now become
\begin{equation}
\partial_t \Sigma + \Psi_z \partial_s \Sigma - \Psi_s \partial_z \Sigma + \Delta c \Psi_s = {\frac {\cal D} {2}} {\bf \nabla}^2 \Sigma, \label{ContinSLb1}
\end{equation}
\begin{equation}
\partial_t {\bf \nabla}^2 \Psi + \Psi_z {\bf \nabla}^2 \Psi_s - \Psi_s {\bf \nabla}^2 \Psi_z = - \lambda \Theta g_0 \beta \partial_s \Sigma + {\frac {1} {R_e}} {\bf \nabla}^2 {\bf \nabla}^2 \Psi. \label{Hydropot1}
\end{equation}
\noindent Using the free boundary conditions
\begin{equation}
\Sigma {\left( 0, 0; t \right)} = \Sigma {\left( 0, 1; t \right)} = 0, \label{BounCondsigma}
\end{equation}
\begin{equation}
\Psi {\left( 0, 0; t \right)} = \Psi {\left( 0, 1; t \right)} = {\bf \nabla}^2 \Psi {\left( 0, 0; t \right)} = {\bf \nabla}^2 \Psi {\left( 0, 1; t \right)} = 0, \label{BounCond}
\end{equation}
\noindent we keep only the lowest harmonic terms in the Fourier expansion of the variables $\Sigma$ and $\Psi$. Note that the characteristic length scale $L_c$ introduced in Section \ref{sec:hydro} has been chosen to be the vertical one.

Using the ansatz
\begin{equation}
\Psi {\left( s, z; \tau \right)} = - {\frac {{\cal D} {\left( 1 + a^2 \right)}} {{\sqrt{2}} a}} {\cal X} {\left( \tau \right)} \sin {\left( \pi a s \right)} \sin {\left( \pi z \right)}, \label{Ansatz1}
\end{equation}
\begin{equation}
\Sigma {\left( s, z; \tau \right)} = B {\left[ {\cal Y} {\left( \tau \right)} \cos {\left( \pi a s \right)} \sin {\left( \pi z \right)} - {\frac {{\cal Z} {\left( \tau \right)}} {\sqrt{2}}} \sin {\left( 2 \pi z \right)} \right]}, \label{Ansatz2}
\end{equation}
\noindent where $a = L_c / L_x$ is the aspect ratio, that is the ratio between the vertical and the horizontal characteristic lengths,
\begin{equation}
B = {\frac {\cal D} {\sqrt{2}}} {\frac {\pi^3 {\left( 1 + a^2 \right)}^3} {\lambda a^2 \Theta g_0 \beta R_e}}, \label{BCoeffic}
\end{equation}
\noindent and neglecting higher harmonic terms, we obtain the Lorenz model \cite{Lorenz}
\begin{equation}
{\dot{\cal X}} = \sigma {\left( {\cal Y} - {\cal X} \right)}, \label{LorenzX}
\end{equation}
\begin{equation}
{\dot{\cal Y}} = {\cal X} {\left( r - {\cal Z} \right)} - {\cal Y}, \label{LorenzY}
\end{equation}
\begin{equation}
{\dot{\cal Z}} = {\cal X} {\cal Y} - b {\cal Z}. \label{LorenzZ}
\end{equation}
\noindent Here, the dot denotes the derivative with respect to the normalized time
\begin{equation}
\tau = {\frac {\pi^2 {\cal D}} {2}} {\left( 1 + a^2 \right)} t. \label{NormalTime}
\end{equation}
\noindent In addition, the control parameters entering the Lorenz equations are given by the expressions
\begin{equation}
\sigma = {\frac {2} {{\cal D} R_e}}, \qquad r = {\frac {{\sqrt{2}} \Delta c} {\pi B}}, \qquad b = {\frac {4} {1 + a^2}}. \label{ControlPar}
\end{equation}

Let us introduce the notation ${\bf W} = {\left( {\cal X}, {\cal Y}, {\cal Z} \right)}$. It is normally assumed that the control parameters $\sigma$, $r$ and $b$ are all positive. In the case, where $r < 1$ the only fixed point of the Lorenz attractor is at the origin ${\bf W}_0 = {\left( 0, 0, 0 \right)}$. This point corresponds to a stable osmotic motion, and all orbits converge to the origin. The latter can be seen easily, since the stability matrix
\begin{equation}
{\widehat{\cal M}_0} = {\left( \begin{array}{ccc}
- \sigma \; & \; \sigma \; & \; 0 \\
r \; & \; -1 \; & \; 0 \\
0 \; & \; 0 \; & \; - b \end{array} \right)}, \label{StabMatrix}
\end{equation}
\noindent possesses eigenvalues
\begin{equation}
\lambda_{1,2} = - {\frac {\sigma + 1} {2}} \pm {\frac {1} {2}} {\sqrt{{\left( \sigma + 1 \right)}^2 - 4 \sigma {\left( 1 - r \right)}}}, \qquad \lambda_3 = - b, \label{EigenValues}
\end{equation}
\noindent which are all negative for $r < 1$.

A pitchfork bifurcation occurs at the value $r = 1$, such that the further growth of $r > 1$ gives rise to two additional critical points
\begin{equation}
{\bf W}_{2,3} = {\left( \pm {\sqrt{b {\left( r - 1 \right)}}}, \pm {\sqrt{b {\left( r - 1 \right)}}}, r - 1 \right)}. \label{CriticalPoint}
\end{equation}
\noindent The stability matrix for the fixed points ${\bf W}_{2,3}$ is
\begin{equation}
{\widehat{\cal M}_1} = {\left( \begin{array}{ccc}
- \sigma \; & \; \sigma \; & \; 0 \\
1 \; & \; -1 \; & \; - d \\
d \; & \; d \; & \; - b \end{array} \right)}, \qquad d = \pm {\sqrt{b {\left( r - 1 \right)}}}. \label{StabMatrix1}
\end{equation}
\noindent The corresponding eigenvalues are the roots of the cubic equation
\begin{equation}
\lambda^3 + {\left( 1 + b + \sigma \right)} \lambda^2 + b {\left( r + \sigma \right)} \lambda + 2 b \sigma {\left( r - 1 \right)} = 0. \label{CubicEquat}
\end{equation}
It is obvious that for $r = 1$ we have three eigenvalues $\lambda_1 = 0$, $\lambda_2 = - b$ and $\lambda_3 = - (1 + \sigma)$, which implies that the osmotic mixing fixed point is marginally stable for $1 < r < r_1$, where
\begin{equation}
r_1 = {\frac {{\left( 1 + b + \sigma \right)}^2} {3b}} - \sigma. \label{Parametr1}
\end{equation}
\noindent For $r_1 < r_c$, where
\begin{equation}
r_c = \sigma {\frac {\sigma + b + 3} {\sigma - b - 1}}, \label{Parametrc}
\end{equation}
\noindent two of the eigenvalues become complex. This means that the system possesses two limit cycles, which are
stable so long as the real part of the complex eigenvalues is negative. For $r = r_c$ these real parts become zero, that is, we have two imaginary eigenvalues. Above $r_c$ the limit cycles become unstable (the complex eigenvalues have positive real parts), and chaos sets in. This implies that $r_c$ is the critical value of the parameter $r$, which characterizes the transition to the turbulent osmotic mixing.

\section{\label{sec:remarks}Concluding Remarks}

In the present paper we offer a simple dynamical model for computing the osmotically driven fluid flow in a variety of complex, non equilibrium situations. Intuitively, the model is appealing, since one can interpret osmotic flows and pressures in terms of a simple mixing of the solvent and solution fluids. Our model can be used for numerical modeling of complex flow patterns in membrane systems, because it provides a more robust approach capable of including many external parameters. 

Using the Oberbeck-Boussinesq approximation, we have derived the basic equations describing the process of forward osmosis, which are very similar to the ones used to model the free Rayleigh-Benard convection. It can be concluded that in some sense the osmotic process is similar to the free convection. The difference is that while in the case of thermal convection the volume expansion is driven by the coefficient of thermal expansion, the key role for the osmotic instability is played by the coefficient of isothermal compressibility $\beta$. In addition, we have shown that the osmotic process represents a propagation of standing waves with time-dependent amplitudes and phase velocity, which equals the current velocity of the solvent passing through the semi-permeable membrane. The evolution of the amplitudes of the osmotic waves is exactly following the dynamics of a strange attractor of Lorenz type with corresponding control parameters.

\begin{acknowledgments}
This research was partially supported by the Bulgarian Science Fund at the Ministry of Education and Science of Bulgaria under contract DTC-02/35.
\end{acknowledgments}

\nocite{*}
\bibliography{aipsamp}% Produces the bibliography via BibTeX.

\begin{thebibliography}{99}
\bibitem{Finkelstein} A. Finkelstein, {\it Water Movement Through Lipid Bilayers, Pores, and Plasma Membranes. Theory
and Reality}, John Wiley and Sons, New York (1987).
\bibitem{Hill} Terrell L. Hill, {\it An Introduction to Statistical Thermodynamics}, Addison-Wesley, Reading Massachusetts (1960).
\bibitem{Reichl} L.E. Reichl, {\it A Modern Course in Statistical Physics}, John Wiley and Sons, New York (2009).
\bibitem{Gruber} M.F. Gruber, C.J. Johnson, C.Y. Tang, M.H. Jensen, L. Yde and C. Hélix-Nielsen, {\it Computational fluid dynamics simulations of flow and concentration polarization in forward osmosis membrane systems}, Journal of Membrane Science {\bf 379}, 488 (2011).
\bibitem{Saltzman} B. Saltzman, {\it Finite Amplitude Free Convection as an Initial Value Problem - I}, Journal of the Atmospheric Sciences, {\bf 19}, 329 (1962).
\bibitem{Klimontovich} Yu.L. Klimontovich, {\it Statistical theory of open systems}, Kluwer Academic Publishers, Dordrecht (1995).
\bibitem{TzenovBook} S.I. Tzenov, {\it Contemporary accelerator physics}, World Scientific, Singapore (2004).
\bibitem{Capillary} S. De Martino, M. Falanga, G. Lauro and S.I. Tzenov, {\it Kinetic derivation of the hydrodynamic equations for capillary fluids}, Physical Review {\bf E 70}, 067301 (2004).
\bibitem{Huang} K. Huang, {\it Statistical Mechanics}, John Wiley \& Sons, New York (1987).
\bibitem{Membrane} T. Kosztolowicz and S. Mrowczynski, "Membrane Boundary Condition", Acta Physica Polonica B, {\bf 32}, 217 (2001).
\bibitem{Lorenz} E.N. Lorenz, {\it Deterministic Nonperiodic Flow}, Journal of the Atmospheric Sciences, {\bf 20}, 130 (1963).
\end{thebibliography}

\end{document}